\documentclass[preprint,superscriptaddress,prb]{revtex4}
\usepackage{graphicx}
\usepackage{epstopdf}

\begin{document}
\title{A graphene Zener-Klein transistor cooled by a hyperbolic substrate}

\date{\today}

\author{Wei Yang}
\affiliation{Laboratoire Pierre Aigrain, D\'epartement de physique de l'ENS, Ecole normale
sup\'erieure, PSL Research University, Universit\'e Paris Diderot, Sorbonne Paris Cit\'e, Sorbonne
Universit\'es, UPMC Univ. Paris 06, CNRS, 75005 Paris, France}
\author{Simon Berthou} \affiliation{Laboratoire Pierre Aigrain, D\'epartement de physique de l'ENS, Ecole normale
sup\'erieure, PSL Research University, Universit\'e Paris Diderot, Sorbonne Paris Cit\'e, Sorbonne
Universit\'es, UPMC Univ. Paris 06, CNRS, 75005 Paris, France}
\author{Xiaobo Lu}
\affiliation{Beijing National Laboratory for Condensed Matter Physics and Institute of Physics,
Chinese Academy of Sciences, Beijing 100190, China}
\author{Quentin Wilmart}\affiliation{Laboratoire Pierre Aigrain, D\'epartement de physique de l'ENS, Ecole normale
sup\'erieure, PSL Research University, Universit\'e Paris Diderot, Sorbonne Paris Cit\'e, Sorbonne
Universit\'es, UPMC Univ. Paris 06, CNRS, 75005 Paris, France}
\author{Anne Denis}
\affiliation{Laboratoire Pierre Aigrain, D\'epartement de physique de l'ENS, Ecole normale
sup\'erieure, PSL Research University, Universit\'e Paris Diderot, Sorbonne Paris Cit\'e, Sorbonne
Universit\'es, UPMC Univ. Paris 06, CNRS, 75005 Paris, France}
\author{Michael Rosticher}
\affiliation{Laboratoire Pierre Aigrain, D\'epartement de physique de l'ENS, Ecole normale
sup\'erieure, PSL Research University, Universit\'e Paris Diderot, Sorbonne Paris Cit\'e, Sorbonne
Universit\'es, UPMC Univ. Paris 06, CNRS, 75005 Paris, France}
\author{Takashi Taniguchi}
\affiliation{Advanced Materials Laboratory, National Institute for Materials Science, Tsukuba,
Japan}
\author{Kenji Watanabe}
\affiliation{Advanced Materials Laboratory, National Institute for Materials Science, Tsukuba,
Japan}
\author{Gwendal F\`eve}
\affiliation{Laboratoire Pierre Aigrain, D\'epartement de physique de l'ENS, Ecole normale
sup\'erieure, PSL Research University, Universit\'e Paris Diderot, Sorbonne Paris Cit\'e, Sorbonne
Universit\'es, UPMC Univ. Paris 06, CNRS, 75005 Paris, France}
\author{Jean-Marc Berroir}
\affiliation{Laboratoire Pierre Aigrain, D\'epartement de physique de l'ENS, Ecole normale
sup\'erieure, PSL Research University, Universit\'e Paris Diderot, Sorbonne Paris Cit\'e, Sorbonne
Universit\'es, UPMC Univ. Paris 06, CNRS, 75005 Paris, France}
\author{Guangyu Zhang}
\affiliation{Beijing National Laboratory for Condensed Matter Physics and Institute of Physics,
Chinese Academy of Sciences, Beijing 100190, China}
\author{Christophe Voisin}
\affiliation{Laboratoire Pierre Aigrain, D\'epartement de physique de l'ENS, Ecole normale
sup\'erieure, PSL Research University, Universit\'e Paris Diderot, Sorbonne Paris Cit\'e, Sorbonne
Universit\'es, UPMC Univ. Paris 06, CNRS, 75005 Paris, France}
\author{Emmanuel Baudin}
\affiliation{Laboratoire Pierre Aigrain, D\'epartement de physique de l'ENS, Ecole normale
sup\'erieure, PSL Research University, Universit\'e Paris Diderot, Sorbonne Paris Cit\'e, Sorbonne
Universit\'es, UPMC Univ. Paris 06, CNRS, 75005 Paris, France}
\author{Bernard Pla\c{c}ais}
\email{bernard.placais@lpa.ens.fr} \affiliation{Laboratoire Pierre Aigrain, D\'epartement de
physique de l'ENS, Ecole normale sup\'erieure, PSL Research University, Universit\'e Paris Diderot,
Sorbonne Paris Cit\'e, Sorbonne Universit\'es, UPMC Univ. Paris 06, CNRS, 75005 Paris, France}

\maketitle \textbf{Engineering of cooling mechanisms is a bottleneck in nanoelectronics. Whereas
thermal exchanges in diffusive graphene are mostly driven by defect assisted acoustic phonon
scattering, the case of high-mobility graphene on hexagonal Boron Nitride (hBN) is radically
different with a prominent contribution of remote phonons from the substrate. A bi-layer graphene
on hBN transistor with local gate is driven in a regime where almost perfect current saturation is
achieved by compensation of the decrease of the carrier density and Zener-Klein tunneling (ZKT) at
high bias. Using noise thermometry, we show that this Zener-Klein tunneling triggers a new cooling
pathway due to the emission of hyperbolic phonon polaritons (HPP) in hBN  by out-of-equilibrium
electron-hole pairs beyond the super-Planckian regime. The combination of ZKT-transport and
HPP-cooling promotes graphene on BN transistors as a valuable nanotechnology for power devices and
RF electronics.}

Energy relaxation in solids is provided by electron-electron interactions and phonon emission. The
former give rise to Wiedemann-Franz (WF) heat conduction to the leads. In diffusive graphene,
acoustic phonon emission is dominated by three-body electron-phonon-impurity supercollisions (SC)
at room temperature \cite{Song2012prl,Graham2013nphys,Betz2013nphys,Laitinen2014nl}. The case of
high-mobility graphene, in spite of its technological interest, has been less investigated. The
suppression of supercollisions and the vanishing of Wiedemann-Franz heat conduction at current
saturation give rise to strongly out-of-equilibrium electron distributions where new cooling
pathways become prominent. Intrinsic optical phonon (OP) cooling is one of those, it was reported
at high density \cite{Mihnev2016ncom} and in suspended graphene \cite{Bo2011nl,Laitinen2014prb}.
Another relaxation mechanism involves interlayer Coulomb coupling in decoupled multilayer epitaxial
graphene \cite{Mihnev2015ncom}. In supported graphene the coupling to remote polar phonons
overwhelms that to OPs \cite{Low2012prb, Viljas2010prb,Barreiro2009prl,DaSilva2010prl}. The case of
hBN supported or encapsulated graphene is emblematic. Firstly, current saturation can be achieved
at low fields
 $\mathcal{E}$  (see Ref.\cite{Meric2008nnanotech}) opening access to the Zener Klein tunneling (ZKT) regime at
high field \cite{Vandecasteele2010prb,Kane2015jpcm}. Secondly, hBN is a uniaxial dielectric that
sustains hyperbolic phonon-polaritons (HPPs)
\cite{Guo2012apl,Biehs2012prl,Biehs2014apl,Dai2015nnano,Kumar2015nl,Giles2016nl,Principi2017prl} in
the two Reststrahlen (RS) bands $\hbar\Omega_I=90$--$100\;\mathrm{meV}$ and
$\hbar\Omega_{II}=170$--$200\;\mathrm{meV}$. As a marked difference with SiO$_2$ surface modes,
HPPs can efficiently radiate energy across the dielectric layer \cite{Biehs2012prl}, avoiding
hot-phonon effects and making an efficient thermal bridge between the graphene channel and the
metallic gate in nanodevices.

By means of sensitive noise thermometry, we show strong evidence for a new and ultra-efficient
cooling pathway beyond the super-Plankian regime that clips the electron temperature when the ZKT
threshold field is reached. This new regime suggests the emission of HPP by out-of-equilibrium
electron-hole pairs. We have investigated  single layer (SLG), bilayer (BLG) and trilayer (TLG)
graphene transistors where similar results are observed (see SI section-III and Fig.S3). However,
we focus here on the BLG sample which is most illustrative essentially due to its nearly energy
independent density of states (DOS) \footnote{Our investigated energy range
($\pm200\;\mathrm{meV}$) excludes contributions from excited subbands and secures the parabolic
band approximation. The gate induced bandgap opening ($\lesssim 20\;\mathrm{meV}$ at the maximum
gate voltage) is null at charge neutrality and small at the scale of the Fermi-energy or the
electronic temperature at large doping.}. In addition, this BLG sample approaches the intrinsic
limit with $v_{sat}\simeq3.10^5\mathrm{m.s^{-1}}$, thereby bringing a more direct insight into the
ultimate relaxation mechanisms where currents and Joule power are maximized.

\section*{Intraband current saturation}

\begin{figure}[hh]
\centerline{\includegraphics[width=14cm]{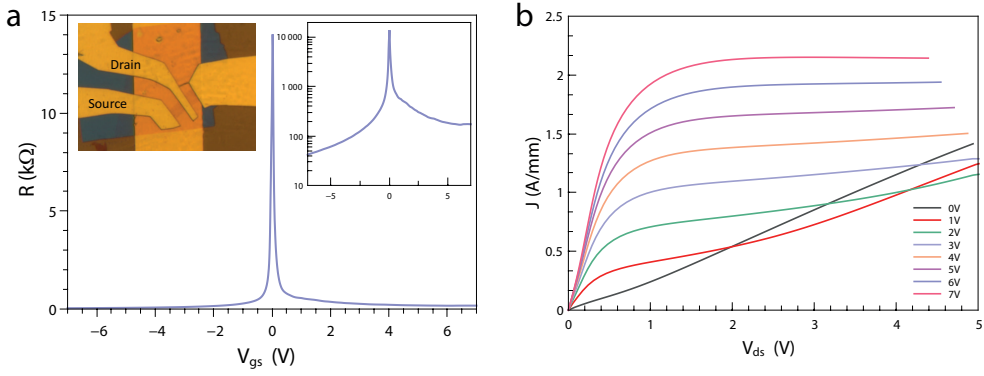}}
\caption{Bottom-gated bilayer graphene on hBN transistor (optical image in panel a-inset).
a) Low-bias transfer curve  $R=1/g_{ds}$ measured at 4.2 Kelvin and $V_{ds}=10$~mV.
A logarithmic plot (inset) shows the small contact resistance in the hole side and a larger one in the electron-side due to contact doping.
Quantum Hall measurements (not shown) allow to deduce the bilayer nature of the sample, its electronic mobility $\mu\simeq3.10^4\mathrm{cm^2V^{-1}s^{-1}}$ and the gate capacitance $C_{g}=1.15\;\mathrm{mF/m^2}$. The gate capacitance mainly arises from the hBN dielectric capacitance
($\epsilon_r\simeq 3.2$) with a negligible serial quantum capacitance correction
$C_Q=e^2DOS=2e^2m^*/(\pi\hbar^2)\simeq40\;\mathrm{mF/m^2}$ (effective mass $m^*\simeq0.03\;m_0$),
which defines the accessible Fermi energy range
$\varepsilon_{F}=\frac{C_g}{C_Q}eV_g=\pm0.2\;\mathrm{eV}$.
b) current saturation for different gate voltages in the electron doped regime (positive bias).}\label{current_saturation.fig1}
\end{figure}

A picture of the BLG transistor and its low-bias resistance are shown in Figure
\ref{current_saturation.fig1}-a. The device is made of a ($L\times W=4\times3\;\mathrm{\mu m}$),
high-mobility  ($\mu\simeq3. 10^4\;\mathrm{cm^2V^{-1}s^{-1}}$) BLG flake exfoliated on a
$23\;\mathrm{nm}$-thick hBN crystal deposited on a metallic (Au) bottom gate, and equipped with
high transparency Pd/Au contacts (see Methods). The gate capacitance
$C_{gs}\simeq1.15\;\mathrm{mF/m^2}$, calibrated against quantum Hall plateaus defines the
accessible Fermi energy range $\varepsilon_{F}=\pm0.2\;\mathrm{eV}$. As seen in
Fig.\ref{current_saturation.fig1}-b, the device demonstrates full current saturation at moderate
and high doping, whereas the emergence of a constant-resistance regime at low doping is the
fingerprint of Zener-Klein tunneling of conductivity $\sigma_{zk}$ \cite{Kane2015jpcm}. As shown in
the Supplementary (section-II and Fig.S2), the full current saturation in
Fig.\ref{current_saturation.fig1}-b results from the balance between the ZKT current and a decrease
of the saturation current by drain doping, which is a property of thin dielectric devices. In the
following we correct for this effect by biasing the BLG sample along constant density lines
($V_{gs}-0.4 V_{ds}=Const$). Although similar results are obtained in the electron doped regime
\footnote{For the sake of clarity, in the following all the qualitative discussions assume electron
doping.}, we focus on the p-doped regime where the contact resistance is minimized (see inset of
Fig.\ref{current_saturation.fig1}-a) \cite{Wilmart2016scirep}.

Fig.\ref{current-noise.fig2}-a  shows the full current-bias characteristics which are consistent
with previous investigations
\cite{Meric2008nnanotech,Freitag2009nl,Barreiro2009prl,DaSilva2010prl,Perebeinos2010prb,Meric2013ieee}.
The length and the higher mobility of our sample make it possible to gain a deeper insight into the
ZKT regime. At low fields, we observe a strong increase of the intraband current with doping (up to
$2\;\mathrm{A/mm}$ at a hole density $p=5. 10^{12}\;\mathrm{cm^{-2}}$). Due to the high mobility,
the current density rapidly reaches important values and saturates. Fig.\ref{current-noise.fig2}-b
shows the differential conductance $\sigma$ at low-field where the ZKT contribution is limited and
intraband transport dominates. It obeys a standard
$\sigma(\mathcal{E})=\sigma(0)/(1+\mathcal{E}/\mathcal{E}_{sat})^2$ dependence
\cite{Meric2008nnanotech}  corresponding to a current density
$J(\mathcal{E})=J_{sat}\mathcal{E}/(\mathcal{E}+\mathcal{E}_{sat})$ where $\mathcal{E}_{sat}$ is
the saturation electric field. From the doping dependence of $\sigma(0)$, we extract a finite field
mobility $\mu\simeq 2.8\;\mathrm{m^2V^{-1}s^{-1}}$ consistent with the zero field measurement
(Fig~\ref{current_saturation.fig1}-a). From  $\mathcal{E}_{sat}$, sketched as a blue dashed line in
Fig.\ref{current-noise.fig2}-b, we define a saturation velocity
$v_{sat}=\mu\mathcal{E}_{sat}=J_{sat}/ne$ and a saturation energy
$\varepsilon_{sat}=\frac{\pi}{2}\hbar k_Fv_{sat}$ \cite{Barreiro2009prl}, which is plotted in
Fig.\ref{current-noise.fig2}-b (inset). At low doping the saturation energy is limited by the Fermi
energy (channel saturation). The saturation of $\varepsilon_{sat}$ at high doping is generally
attributed to OP (or remote phonon) scattering (energy $\hbar\Omega$), according to
$\varepsilon_{sat}=\hbar \Omega$.\cite{Barreiro2009prl} Both the linear increase
$\varepsilon_{sat}\propto | \varepsilon_{F}|$ and the trend to saturation are observed. We extract
the asymptotic limit by fitting the data to the empirical formula
$\varepsilon_{sat}=(\varepsilon_{F}^{-2}+(\hbar\Omega_{sat})^{-2})^{-1/2}$. We deduce
$\hbar\Omega_{sat}\simeq95\pm5\;\mathrm{meV}$ which is consistent with remote phonons $\hbar
\Omega_I$ of the lower RS band of hBN.

\section*{The Zener-Klein tunneling regime}

At higher bias, interband (ZKT) transport takes over, leading to a constant differential
conductance $\sigma_{zk}\simeq1\;\mathrm{mS}$. Similar behavior is observed in SLG and TLG devices
(SI)  with $\sigma_{zk}^{SLG}\simeq1.2\;\mathrm{mS}$ and $\sigma_{zk}^{TLG}\simeq2\;\mathrm{mS}$.
In analogy with Klein tunneling in abrupt p-n junctions in graphene \cite{Katsnelson2006nphys}, ZKT
is constrained by energy/momentum conservation \cite{Kane2015jpcm} which sets a threshold field at
$\mathcal{E}_{zk}=2\varepsilon_F/(el_{zk})$, where $l_{zk}$ is the doping-dependent tunneling
length. ZKT bears strong analogies with optical pumping as it involves vertical interband
transitions (bound by Pauli blocking) with electron-hole pair creation at a rate
$\dot{n}_{e-h}=ek_F/(\hbar\pi^2)(\mathcal{E}-\mathcal{E}_{zk})$. As a marked difference, the
pumping energy window increases linearly with the applied field $\mathcal{E}-\mathcal{E}_{zk}$. In
the absence of theoretical prediction for BLG-ZKT,  we rely on the transmission of a sharp BLG p-n
junction \cite{Katsnelson2006nphys}, $D=k_FW/4\pi$ to deduce the ZKT conductance. To account for
the finite length of the junction, we introduce a transparency factor $\alpha_{zk} \simeq 0.3$
(deduced from the noise measurements, see below). Thus, the ZKT conductance reads
$\sigma_{zk}=\alpha_{zk}\;\frac{4e^2}{h}\;(k_Fl_{zk})/4\pi$. In this simple picture, a doping and
field independent ZKT conductivity translates into a constant $\alpha_{zk}k_Fl_{zk}$ that we deduce
from the low-doping data where the ZKT regime is prominent. We are thus able to compute the
threshold field $\mathcal{E}_{zk}=2|\varepsilon_F|/(el_{zk})$ for each carrier density (red dashed
line in Fig.\ref{current-noise.fig2}-a). In the investigated carrier density range we find
$l_{zk}\propto k_F^{-1} \gtrsim0.8\;\mathrm{\mu m}$ (at $n=5.10^{12}\;\mathrm{cm^{-2}}$) which is
significantly smaller than sample length $L$. The main outcome of our current-bias analysis is that
in quasi intrinsic samples intraband current saturation and Zener-Klein tunneling have different
onset fields (blue and red lines in Fig.~\ref{current-noise.fig2}-b), especially at large doping,
$|\varepsilon_F|=200\;\mathrm{meV}$, where $\mathcal{E}_{sat}\simeq 90\;\mathrm{mV/\mu m}$ and
$\mathcal{E}_{zk}\simeq 500\;\mathrm{mV/\mu m}$.

\section*{Noise thermometry}

\begin{figure}[hh]
\centerline{\includegraphics[width=12cm]{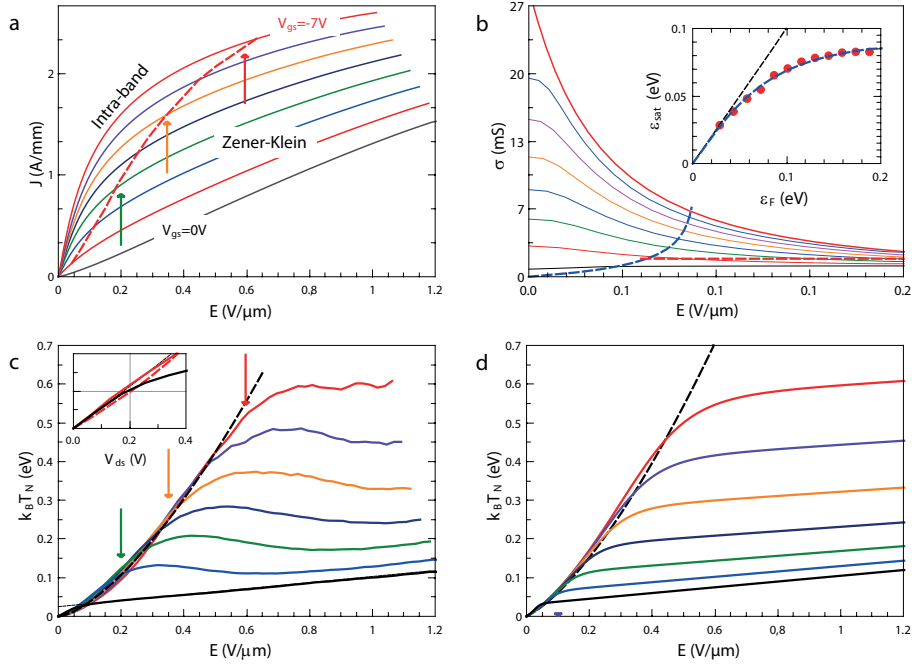}}
\caption{
 \textbf{a)}: Non-linear current-field characteristics of the BLG on hBN transistor in the hole doped regime. The gate voltage (and hence carrier density and Fermi energy) increases linearly in the range $V_{gs}=[-0,-7]\;\mathrm{V}$, $p=[0,5]\times10^{12}\;\mathrm{cm^{-2}}$ and $|\varepsilon_{F}|=[0,0.2]\;\mathrm{eV}$. The onset field for Zener-Klein tunneling $\mathcal{E}_{zk}$ (see main text) is shown as a red dashed line in the plot.
 \textbf{b)}:  Low field dependence of the differential conductivity. Blue dashed line : saturation field $\mathcal{E}_{sat}$, corresponding to  $J=J_{sat}/2$ and $\sigma(\mathcal{E})=\sigma(0)/4$. Note that the ZKT onset field (red dashed line) fulfills $\mathcal{E}_{zk}/\mathcal{E}_{sat} > 1$.
Inset : Fermi energy dependence of the saturation energy $\varepsilon_{sat}$ (defined in the text) and its fitting to $\varepsilon_{sat}=(\varepsilon_{F}^{-2}+(\hbar\Omega_{sat})^{-2})^{-1/2}$ with $\hbar\Omega_{sat}=95\pm 5\;\mathrm{meV}$.
 \textbf{c)}: Bias field dependence of the noise temperature $T_N$. Two regimes are observed, a steep increase at low field followed by a quasi-saturation above a doping dependent threshold $V_{on}=\mathcal{E}_{on}L$. At zero doping $V_{on}\simeq 0.2\;\mathrm{V}\simeq\hbar\Omega_{II}$, the HPP phonon energy (inset).
 \textbf{d)}: Calculated $k_BT_N(V_{ds})$ plots using the heuristic model described in the text,
 including WF cooling at low field and out-of-equilibrium HPP emission at high field.}\label{current-noise.fig2}
\end{figure}

Noise thermometry, combined with Joule heating, is a powerful tool to investigate energy relaxation
\cite{Wu2007prl,Chaste2010apl,Santavicca2010nl,Voisin2015jpcm}. It relies on the measurement of
current noise power density $S_{I}(f)$ at a frequency large enough to exceed the $1/f$ noise corner
frequency which increases with bias. The noise temperature (equal to the electron temperature in
quasi-equilibrium situations) is experimentally defined as $k_BT_N= S_{I} (L/4 \sigma W)$.
Technically we have adapted our standard setup \cite{Chaste2010apl,Betz2012prl,Brunel2015jpcm} to
work in the $1$--$10\;\mathrm{GHz}$ band (see SI section-I and Fig.S1) and accommodate the high
bias conditions. Previous studies have investigated the thermalization sequence in graphene samples
upon increasing power. This sequence usually starts with WF heat conduction on a limited window in
diffusive samples and then displays the signatures of various electron-AC phonon coupling
mechanisms \cite{Betz2012prl,Fong2013prx,McKitterick2016prb} and the emergence of new mechanisms
like SCs \cite{Betz2013nphys,Laitinen2014nl} or interaction effects with Dirac fluid behavior at
low density \cite{Crossno2016science}. The bottom line of these studies is the observation of
power-laws $P_{cool}\propto T_e^\beta$ between cooling power $P_{cool}$ and electronic temperature
$T_e$. Fig.\ref{current-noise.fig2}-c shows the very peculiar thermal behavior of high-mobility
graphene on hBN at high power ($P_{heat}\lesssim 2\;\mathrm{GW m^{-2}}$), with an abrupt switching
between two cooling mechanisms and the clipping of the electronic temperature at high bias.

The low-bias mechanism is naturally WF-cooling that develops on a wide window due to the large heat
conductivity $\kappa\propto\sigma$ and the absence of SCs in high-mobility graphene. It relies on
solutions of the (1D) heat equation  $\frac{1}{2}\mathcal{L}\sigma \partial^2 T^2/\partial
x^2=P_J$, where $P_J$ is the Joule heating density and $\mathcal{L}= (\pi^2 k_B^2)/(3e^2)$ is the
Lorenz number. Assuming uniform Joule heating and accounting for cold contacts, one gets
$k_BT_N=<k_BT_e>=\mathcal{F}Le\mathcal{E}$ with the Fano factor $\mathcal{F}=\frac{\sqrt{3}}{8}$.
Experimentally, the temperature  shows a superlinear $T_N(V_{ds})$ behavior
(Fig.\ref{current-noise.fig2}-c) that can be explained by the current saturation discussed above
and the related increase of the differential conductance $\sigma(\mathcal{E})$
(Fig.\ref{current-noise.fig2}-b). The $T_N\propto\sqrt{\mathcal{E} \cdot J/\sigma}$ scaling in
Fig.\ref{thermal.fig3}-b confirms the WF nature of cooling at low bias. Taking $\mathcal{F}=0.1$
(see below) and $\mathcal{E}_{sat}=90\;\mathrm{mV/\mu m}$ we can reproduce the superlinear law with
$k_BT_N=\mathcal{F}L\mathcal{E}\sqrt{1+\mathcal{E}/\mathcal{E}_{sat}}$
(Fig.\ref{current-noise.fig2}-c,d dotted lines). Note that $T_N\propto L$ in the WF regime so that
smaller temperatures would be observed in shorter samples.

As seen in Fig.\ref{current-noise.fig2}-c the noise temperature deviates from the WF cooling limit
above a doping dependent onset voltage $V_{on}=L\mathcal{E}_{zk}$ and saturates at large electric
field (and Joule power) indicating a very efficient cooling mechanism. A crucial difference between
the low-bias and high-bias cooling mechanisms lies in their opposite dependence on carrier
concentration as shown in Fig.~\ref{thermal.fig3}-a. The usual observation -the larger the carrier
density, the lower the electron temperature- which holds for most cooling mechanisms reported so
far (including ACs, SCs, OPs, or WF \cite{Betz2013nphys, Bistritzer2009}) breaks down for the
high-bias regime. The existence of plateaus, and the increase of the plateau-temperature with
doping, point to the onset of a new cooling mechanism driven by Pauli blocking (see arrows in
Fig.\ref{current-noise.fig2}-a and c). A second hint on this cooling process is given by the lower
limit  $V_{on}\sim0.2\;\mathrm{V}$ at neutrality (inset). This feature, also seen in the SLG and
TLG noise data (see SI section-III and Fig.S3), points to an activation energy close to the second
RS band of BN, $\hbar\Omega_{II}\simeq0.2\;\mathrm{eV}$.

Let us first discuss the intrinsic OP relaxation cooling pathway that has been reported in carbon
nanotubes \cite{Yao2000prl,Bourlon2004prl} and graphene \cite{Barreiro2009prl,Laitinen2014prb}. The
question arises especially as the OP energy window ($\hbar\Omega_{OP}=170$--$200\;\mathrm{meV}$) is
 comparable with that of type-II HPPs and can a priori also explain the voltage threshold $V_{on}\sim0.2\;\mathrm{V}$ at neutrality.
 The mechanisms are different: instead of a
Fr\"ohlich coupling for substrate polar phonons (SPhPs or HPPs), non-polar OPs are coupled to
electrons via the deformation potential giving rise to a smaller relaxation rate. According to
theory for thermal emission \cite{Low2012prb}, the OP cooling power increases with temperature and
doping, which is at variance with our observations that cooling increases at temperature saturation
(Fig.\ref{current-noise.fig2}-c) or drop down (see SI section-IV and Fig.S4). Similarly, the OP
cooling power increases with doping \cite{Low2012prb}, which is in conflict with the observed rise
in temperature with doping in the ZKT regime at a given joule power (Fig\ref{thermal.fig3}-a). To
our knowledge there is no theory for non-thermal emission such as that involved in the Zener-Klein
regime. To settle this issue further, we have performed an in-situ Raman spectroscopy diagnosis of
the OP occupation number. It is based on monitoring the Stokes/anti-Stokes G-peaks ratio amplitude
as function of bias (see SI section-V and Fig.S5); the anti-Stokes 2D-peak escape detection and we
assume, following Ref.\cite{Laitinen2014prb}, an equal contribution of zone edge and zone center OP
cooling. We detect a finite OP population at high electronic temperature, but it is at least 4
times too small to explain the large cooling powers of our experiments.

\section*{Cooling by hyperbolic BN phonon polaritons}

In isotropic polar materials, OPs are responsible for a RestStrahlen band (RS) in which light
propagation is forbidden. Nevertheless, surface modes polaritons (SPhPs) can develop which create a
near-field in the vicinity of the interfaces, to which the electrons of graphene can efficiently
couple. The strong unixial character of hBN is responsible for the splitting of the RS band into a
lower out-of-plane band (90-100 meV) and a higher in-plane band (170-200 meV). In these RS bands,
where each mode brings a dielectric function with real parts of opposite signs, the usual
evanescent SPhPs are replaced by propagating hyperbolic phonon polariton (HPP) modes
\cite{Kumar2015nl}. The large number of HPP modes (the number of branches equals the number of BN
layers) strongly enhances the cooling capability of HPPs compared to SPhPs. Furthermore, while the
vertical transport of energy by SPhPs is limited to their evanescent decay length (about
$\sim1\;\mathrm{nm}$ in our situation), energy transport by HPPs is only limited by their
anharmonic decay which leads to characteristic depths of $\sim30\;\mathrm{nm}$. In our device, the
efficient coupling to the hBN layer opens up thermal pathways reaching the gold backgate where heat
is efficiently drained away from the transistor (inset of Fig.~\ref{fig:superplanck}-a).

In this context, heat transfer from the graphene layer to the hBN substrate can be seen as
black-body radiation into a material bearing specific hyperbolic modes. In vacuum, black-body
radiation in the far field is strongly constrained by the light cone ($k_\parallel \le
k_0=\omega/c$) that puts severe restrictions on the momentum exchange. This picture has to be
revisited when the distance $d$ between the black-body and the dielectric becomes shorter than the
thermal wavelength $\lambda_T=\hbar c/k_B T$, reaching the so-called super-Planckian regime where
thermal coupling mainly occurs through evanescent modes \cite{Biehs2010} up to a wavevector $1/d$.
The thermal contact is reached when $d \lesssim \lambda_F$ as the momentum exchange becomes limited
by the electron wavevectors.

Real bodies are characterized by their emissivity $M(\omega)$, that is their relative radiative
efficiency compared to the black-body emission. Fig.~\ref{fig:superplanck}-a represents the
experimental average emissivity $M$ of our device as a function of the temperature. This emissivity
is the ratio of the Joule power to the maximum theoretical super-Planckian radiation power of the
BLG on the hBN slab (see SI section-VI). If super-Planckian radiation was the main thermal channel,
a smoothly decreasing emissivity would be observed as calculated in Fig.~\ref{fig:superplanck}-b.
In contrast, experimental data show a strong deviation for both the low and high temperature sides.
The apparently diverging emissivity for $k_BT\lesssim40\;\mathrm{meV}$ is an artefact due to the
increasing contribution of the WF cooling at low $T$. Above $k_BT=90\;\mathrm{meV}$, the
super-Planckian HPP emissivity becomes significant and scales similarly to WF cooling. This
contribution of HPP cooling in the intermediate temperature regime ($40\;\mathrm{meV}< k_B T <
E_F$) accounts for the reduced Fano factor $\mathcal{F} \simeq 0.1$ reported consistently in hBN
supported graphene samples.

The most striking feature in Fig.~\ref{fig:superplanck}-a is a sudden jump of the emissivity at a
doping dependent temperature threshold. This behavior cannot be understood within a thermal scheme
since the emissivity at fixed doping should only show a smooth dependence with temperature. This
dramatic increase of the emissivity by more than a decade shows that an ultra efficient, strongly
out-of-equilibrium process sets in. Interestingly, this new thermal channel arises concomitantly
with the switch of the transistor in the ZKT regime (Fig.~\ref{current-noise.fig2}-a and b) that is
equivalent to an electrical pumping of electron-hole pairs. Thus, we suggest that this thermal
channel is due to the emission of HPPs from an inverted electron-hole pair population. In this
respect, the threshold voltage ($V_{ds}\simeq 0.2\;\mathrm{V}$) near neutrality shows that e-h
pairs generated by ZKT with an energy below $\hbar \Omega_{II}$ are naturally unable to cool the
sample. In total, noise thermometry allows to conclude that beyond a first regime which is most
probably thermal, HPP emission above the ZKT onset field is a highly out-of-equilibrium process.

The temperature plateaus observed in Fig\ref{current-noise.fig2}-c at high doping when cranking up
the bias above the ZKT threshold show that the out-of-equilibrium emission of HPPs yields a cooling
power $P_{HPP}$ that can compensate the excess Joule power $\Delta P_J$: $\Delta P_J=P_{HPP}$. In
the saturation regime and neglecting the ZKT current with respect to the intraband current, the
excess Joule power reads $\Delta P_J \simeq J_{sat}(\mathcal{E}-\mathcal{E}_{zk})=2
\varepsilon_{sat} ek_F/(\pi^2 \hbar) \times (\mathcal{E}-\mathcal{E}_{zk})$, whereas the power
drained away by HPP emission reads $P_{HPP}= \dot{n}_{e-h} \hbar \Omega_{II}$ with, for e-h pairs
created by ZKT, $\dot{n}_{e-h}=ek_F/(\hbar\pi^2)(\mathcal{E}-\mathcal{E_{zk}})$. Obviously, those
powers equilibrate provided that $2 \varepsilon_{sat}\simeq \hbar \Omega_{II}$, which is roughly
the case in hBN because $\varepsilon_{sat} \simeq \hbar \Omega_I \simeq \hbar \Omega_{II}/2$.
Interestingly, this observation shows that in quasi-intrinsic samples, the temperature saturation
at high doping ultimately arises from the peculiar frequencies of the hBN RS bands. Taking this
effect into account together with the nonlinear WF cooling described above, we have plotted in
Fig.\ref{current-noise.fig2}-d a simulation of the noise temperature as a function of bias voltage
using $\alpha_{zk}=0.3$ as a free parameters and $\mathcal{F}=0.1$ from the WF scaling in
Fig.\ref{thermal.fig3}-b. The agreement with experiment is good and supports our heuristic model,
in particular our main hypothesis that HPP cooling can fully compensate Joule heating so that
electronic temperature itself is clipped. Furthermore, we anticipate that in diffusive samples, the
lower saturation current (and thus lower Joule power) would lead to a lower equilibrium
temperature. Actually this is observed in a second BLG device having a thicker ($200\;\mathrm{nm}$)
hBN dielectric (see SI section-IV and Fig.S4), where a smaller Joule power due to a smaller
$\sigma_{zk}$ leads to an imbalance of Joule heating and HPP cooling and a drop down of the noise
temperature in the ZKT regime.

In order to characterize further the non-equilibrium HPP emission, it is enlightening to estimate
the steady density of electron-hole pairs in the ZKT regime. In fact, for non-thermal electron
distributions, the noise temperature $T_N$ has an additional contribution above $T_e$ that is
directly related to the presence of non-equilibrium holes. Owing to the constant DOS of BLG, this
correction can be captured by splitting the noise temperature integral along the conduction and
valence bands and writing, for an electron doped BLG, $k_BT_N\simeq k_BT_e+n_{e-h}/DOS$ with
$k_BT_e\simeq \int_{0}^\infty{f(1-f)dE}$  and $n_{e-h}=\int_{-\infty}^0{DOS(1-f)dE}$ as
$f(E<0)\lesssim1$ in weak ZK tunneling conditions. This correction sets the absolute noise floor
for a cold BLG ZKT transistor at  $k_BT_N= 2 n_{e-h}/DOS$. In the steady state, the recombination
of e-h pairs into HPPs at a rate $\dot{n}_{e-h}= -n_{e-h}/\tau$ (where $\tau$ is an effective HPP
emission time) just equilibrates the e-h generation by ZKT. We thus obtain
$n_{e-h}=2\tau\sigma_{zk}/el_{zk}\times (\mathcal{E}-\mathcal{E}_{zk})$ consistent with the
residual linear dependence $T_N(\mathcal{E})$ observed in the low-doping limit where the ZKT regime
is most developed (Fig.~\ref{current-noise.fig2}-c). The slope of $T_N(\mathcal{E})$ near
neutrality yields  $\tau \simeq 0.5\;\mathrm{ps}$, approaching the minimum emission time
$\simeq0.13\;\mathrm{ps}$ (see SI section-VI). Note that the measured  time $\tau$ is much larger
that the intraband electron-electron relaxation time ($\simeq
50\;\mathrm{fs}$)\cite{Brida2013ncomm,Kadi2015scirep} but roughly $4$ times smaller than the
intrinsic OP emission time entering the cooling rate \cite{Low2012prb}. This analysis provides a
consistent picture of the cooling pathway where electrons in the conduction band rapidly thermalize
with the Fermi sea, whereas intrinsic energy relaxation by OPs is quenched by a faster coupling to
HPPs. We note that fast HPP relaxation has been recently reported in a photo-thermoelectric
photovoltage experiment \cite{Tielrooij2017arXiv}.

\section*{Conclusion}

In conclusion, using combined transport and noise thermometry we have shown that quasi-intrinsic
bilayer graphene on hBN transistors have remarkable thermal properties, dominated by
Wiedemann-Franz conduction and hBN hyperbolic phonon polariton emission. In particular, we have
unveiled a new out-of-equilibrium HPP emission process subsequent to the generation of
electron-hole pairs by Zener-Klein tunneling, which yields to the temperature plateaus observed at
high doping. A direct signature of this non-equilibrium hole population is observable as a linear
correction to the electronic temperature in the noise power in the low-doping limit from which we
estimate an HPP relaxion time $\simeq0.5\;\mathrm{ps}$. This GoBN technology based on local gating
of high mobility graphene through a thin hBN layer opens up many perspectives : in terms of
applications it makes up a promising platform for RF power amplification and for the design of
original cooling pathways in nano-devices; in terms of basic science it opens up the study of cold
cooling pathways involving out-of-equilibrium carrier generated by tunneling processes, and
promotes graphene as a dedicated source for HPP optics.

\section*{Methods}

The graphene boron nitride heterostructures (including monolayer, bilayer, and trilayer graphene)
are assembled by the dry transfer technique \cite{Dean2010nnano} and the devices are fabricated by
e-beam lithography. We first deposit the bottom gate (width $15\;\mathrm{\mu m}$) and coplanar
waveguide using $2/50\;\mathrm{nm}$ Cr/Au metallization on a high resistivity Si substrate covered
by a $285\;\mathrm{nm}$ SiO$_2$. A high quality hBN crystal is then stamped on top of the gate
under microscope. The PDMS stamp is removed with acetone and the sample is further cleaned by a
$1-2\;\mathrm{hours}$ annealing at a temperature of  $300-400\;\mathrm{^{\circ}C}$ under
$200\;\mathrm{sccm}$-Ar / $50\;\mathrm{sccm}$-H$_2$ flow. The exfoliated graphene flakes are
transferred on hBN using a PPC stamp, removed again with acetone and annealed as before. Graphene
remains as exfoliated to avoid degrading mobility etching processes. Finally $50/50\;\mathrm{nm}$
Pd/Au source and drain contacts are deposited. Our devices are uncapped to secure low contact
resistance at high frequency and bias. Chemical surface contamination is removed in situ with a
current annealing at low temperature.

The data that support the plots within this paper and other findings of this study are available
from the corresponding author upon reasonable request

\textbf{Author contributions} WY, EB, CV and BP conceived the experiment and developed the models.
WY, SB conducted the measurements. AD  designed the sample holder.  WY, XL, MR, TT, KW, GZ
participated to sample fabrication. WY, SB, GF, JMB, EB, CV and BP participated to the data
analysis.  WY, EB, CV and BP wrote the manuscript with contributions from the
coauthors.

\textbf{Data availability statement}
The data that support the plots within this paper and other
findings of this study are available from the corresponding author upon reasonable request.

\textbf{Additional information} Competing financial interests: The authors declare no competing
financial interests. Supplementary information is available in the online version of the paper.
Reprints and permission information is available online at www.nature.com/reprints. Correspondence
and requests for materials should be addressed to [BP]

\begin{acknowledgments}
The research leading to these results have received partial funding from the European union under
grant N:696656 Graphene Flagship, and from the french ANR under the grant ANR-14-CE08-018-05
"GoBN". G.Z. acknowledges the financial supports from the National Basic Research Program of China
(973 Program) under the grant No:2013CB934500, the National Science Foundation of China (NSFC)
under the grant No:61325021.
\end{acknowledgments}

\newpage

 \begin{figure}[hh]
\centerline{\includegraphics[width=15cm]{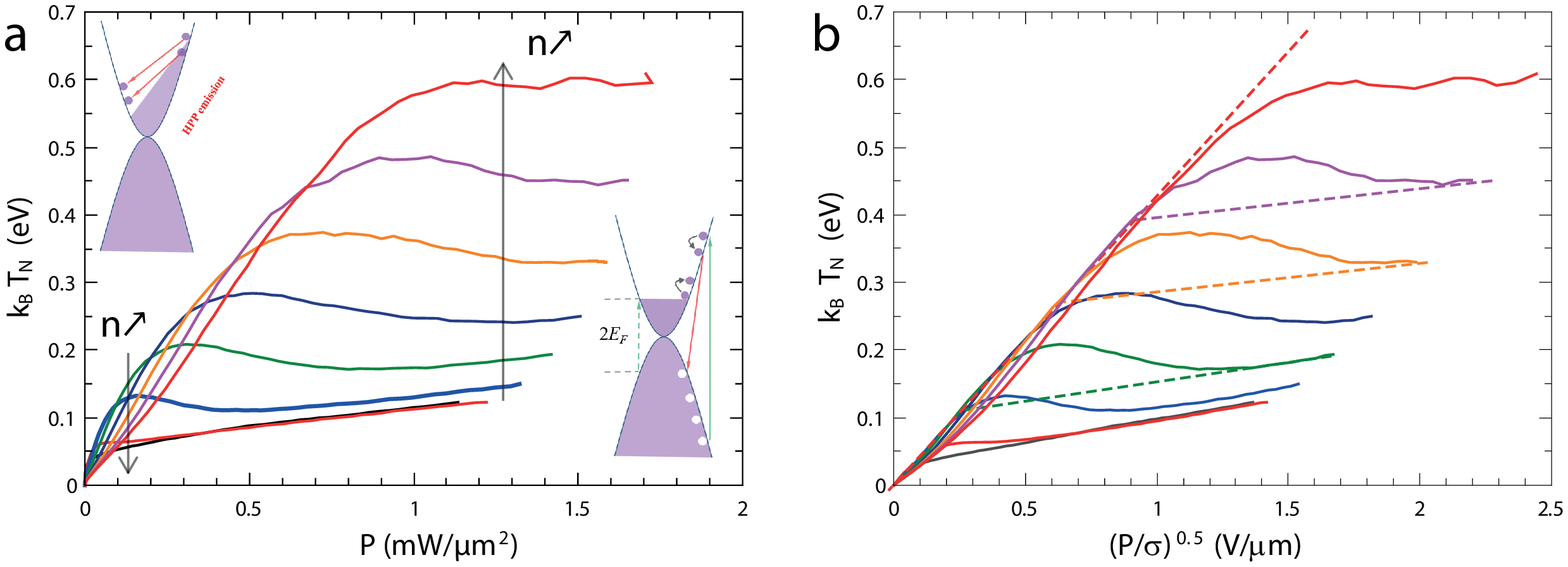}}
\caption{ (a) Noise temperature as a function of the Joule heating. The sketches represent the dominating cooling processes :
intra-band electron-electron interactions at low field and interband HPP emission at high field. (b) Wiedemann-Franz scaling
of the noise temperature data; from the subthreshold slope we deduce the Fano factor $\mathcal{F}\simeq0.105$ and a residual
slope $\mathcal{F}\simeq0.015$ above the threshold.}\label{thermal.fig3}
\end{figure}

 \begin{figure}[h]
 \begin{center}
 \centerline{\includegraphics[width=15cm]{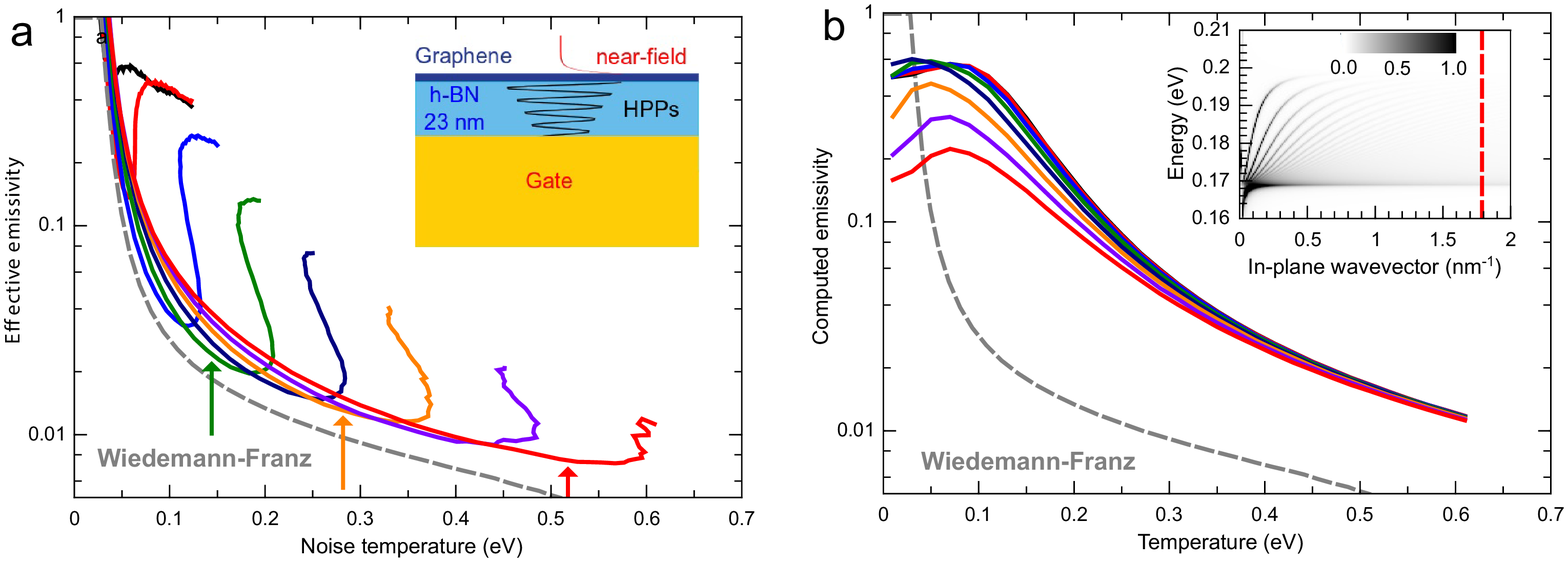}}
\caption{ (a) Experimental emissivity computed as the ratio of the Joule power to the theoretical super-Planckian power.
Inset : Sketch of the radiative heat transfer by HPPs. (b) Theoretical emissivity of
the BLG-hBN(23 nm)-Au stack computed as a function of the temperature of the graphene electrons for a BLG in the local conductivity approximation.
Inset : Monochromatic emissivity $M(\omega,k)$ of the BLG/hBN/Au stack, showing strong Fabry-Perot-like resonances in the 23~nm-thick
hBN layer. Dashed line : Wavevector cut-off $k_c(\omega, E_F,T)$ used for the calculation of the super-Planckian radiated power.
The emissivity of the main panel is the frequency and wavevector average of $M(\omega,k)$ over the whole RS band.}\label{fig:superplanck}
 \end{center}
\end{figure}


\begin{thebibliography}{}


\bibitem{Song2012prl} Song, J. C. W., Reizer, M. Y.  and Levitov,  L. S.,  \emph{Phys. Rev. Lett.}
    \textbf{109},
    106602 (2012). \textit{Disorder-assisted electron-phonon scattering and cooling pathways in
    graphene.}

\bibitem{Graham2013nphys} Graham, M., Shi, S.-F.,  Ralph, D. C.,  Park, J.  and McEuen,  P. L.,
    \emph{Nat. Phys.} \textbf{9}, 103 (2013). \textit{Photocurrent measurements of supercollision cooling in
    graphene}

\bibitem{Betz2013nphys} Betz, A.C. , et al., \emph{Nat. Phys.} \textbf{9}, 109 (2013).
 \textit{Supercollision cooling in undoped graphene.}

\bibitem{Laitinen2014nl} Laitinen, A. , et al., \emph{Nano Lett.} \textbf{14}, 3009 (2014).
    \textit{Electron phonon coupling in suspended graphene: supercollisions by ripples.}

 \bibitem{Mihnev2016ncom} Mihnev, M.T., et al., \emph{Nat. Comm.} \textbf{7}, 11617 (2016).
 \textit{Microscopic origins of the terahertz carrier relaxation and cooling dynamics in graphene.}

\bibitem{Bo2011nl} Gao, B., et al., \emph{Nano Lett.} \textbf{11}, 3184 (2011).
    \textit{Studies of Intrinsic Hot Phonon Dynamics in Suspended Graphene by Transient Absorption Microscopy}

 \bibitem{Laitinen2014prb} Laitinen, A., et al., \emph{Phys. Rev. B} \textbf{91}, 121414 (R) (2014).
\textit{Electron-optical phonon coupling in suspended graphene bilayer.}

\bibitem{Mihnev2015ncom} Mihnev, M.T., et al., \emph{Nat. Comm.} \textbf{6}, 8105 (2015).
 \textit{Electronic cooling via interlayer Coulomb coupling in multilayer epitaxial graphene.}

\bibitem{Low2012prb} Low, T., Perebeinos, V.,  Kim, R., Freitag,  M.,  and P. Avouris, P.,
  \emph{Phys. Rev. B} \textbf{86}, 045413 (2012).
 \textit{Cooling of photoexcited carriers in graphene by internal and substrate phonons}.

\bibitem{Viljas2010prb} Viljas, J. K., Heikkil\"{a}, T.T., \emph{Phys. Rev. B} \textbf{81}, 245404
    (2010). \textit{Electron-phonon heat transfer in monolayer and bilayer graphene}.

\bibitem{Barreiro2009prl} Barreiro, A., Lazzeri, M.,  Moser, J., Mauri, F.,  Bachtold, A.,
    \emph{Phys.  Rev. Lett.} \textbf{103}, 076601 (2009).
    \textit{Transport properties of graphene in the high-current limit}

\bibitem{DaSilva2010prl} DaSilva, A.M.,  Zou, K., Jain, J.K., Zhu,  J.,  \emph{Phys. Rev. Lett.}
    \textbf{104}, 236601 (2010).
    \textit{Mechanism for current saturation and energy dissipation in graphenet transistors}

\bibitem{Meric2008nnanotech} Meric, M., et al., \emph{Nat. Nanotech.} \textbf{3}, 654 (2008).
 \textit{Current saturation in zero-bandgap, topgated graphene field-effect transistors}

\bibitem{Vandecasteele2010prb} Vandecasteele, N.,  Barreiro, A., Lazzeri, M., Bachtold, A., Mauri,
    F.,     \emph{Phys. Rev. B} \textbf{82}, 045416 (2010).
    \textit{Current-voltage characteristics of graphene devices: Interplay between Zener-Klein tunneling and defects}

\bibitem{Kane2015jpcm} Kane, G., Lazzeri, M., Mauri, F., \emph{J. Phys.: Condens. Matter}
    \textbf{27},  164205 (2015).
    \textit{High-field transport in graphene: the impact of Zener tunneling}

\bibitem{Guo2012apl}  Guo, Y., Cortes, C.L., Molesky, S., Jacob, Z., \emph{App. Phys.  Lett.}
    \textbf{101},131106 (2012)
    \textit{Broadband super-Planckian thermal emission from hyperbolic metamaterials}

\bibitem{Biehs2012prl}  Biehs, S.-A., Tschikin,  M.,  Ben-Abdallah, P.,  \emph{Phys. Rev. Lett.}
    \textbf{109}, 104301 (2012)
 \textit{Hyperbolic metamaterials as an analog of a blackbody in the near field}

\bibitem{Biehs2014apl} Biehs, S.-A., Tschikin, M., Messina, R.,  Ben-Abdallah, P., \emph{App. Phys.
 Lett.} \textbf{105},161902 (2014).
 \textit{Super-Planckian far-zone thermal emission from asymmetric hyperbolic metamaterials}

\bibitem{Dai2015nnano} Dai, S., et al.,  \emph{Nat. Nanotech.} \textbf{10}, 682 (2015).
    \textit{Graphene on hexagonal boron nitride as a tunable hyperbolic metamaterial}

\bibitem{Kumar2015nl} Kumar, A.,  Low, T., Fung, K. H.,  Avouris, P.,  Fang, N. X., \emph{Nano
    Lett.} \textbf{15}, 3172 (2015).
    \textit{Tunable light-matter interaction and the role of hyperbolicity in graphene-hBN system}

\bibitem{Giles2016nl} Giles, A.J., et al., \emph{Nano Lett.} \textbf{16}, 3858 (2016)
    \textit{Imaging of anomalous internal reflections of hyperbolic phonon-polaritons in hexagonal boron nitride}

 \bibitem{Principi2017prl} Principi, A., et al., \emph{Phys. Rev. Lett.} \textbf{118}, 126804 (2017)
 \textit{Super-Planckian electron cooling in a van der Waals stack}

\bibitem{Wilmart2016scirep} Wilmart, Q., et al., \emph{Scientific Reports} \textbf{6}, 21085
    (2016).
    \textit{Contact gating at GHz frequency in graphene}

\bibitem{Freitag2009nl} Freitag, M., et al., \emph{Nano Lett.} \textbf{9}, 1883 (2009).
    \textit{Energy dissipation in graphene field-effect transistors}

\bibitem{Perebeinos2010prb} Perebeinos, V., Avouris, P., \emph{Phys. Rev. B} \textbf{81}, 195442
    (2010).
    \textit{Inelastic scattering and current saturation in graphene}

\bibitem{Meric2013ieee} Meric, N., et al., \emph{Proc. IEEE} \textbf{101}, 1609 (2013).
    \textit{Graphene field-effect transistors based on boron–nitride dielectrics}

\bibitem{Katsnelson2006nphys} Katsnelson, M. I.,  Novoselov, K. S., Geim, A. K. \emph{Nat. Phys.}
    \textbf{2}, 620 (2006).
    \textit{Chiral tunnelling and the Klein paradox in graphene}

\bibitem{Wu2007prl} Wu, F., et al. \emph{Phys. Rev. Lett.} \textbf{99}, 156803 (2007). \textit{Shot
    noise with interaction effects in single-walled carbon nanotubes}

\bibitem{Chaste2010apl} Chaste, J., et al., \emph{Appl. Phys. Lett.} \textbf{96}, 192103 (2010).
    \textit{Thermal shot noise in top-gated single carbon nanotube field effect transistors}

\bibitem{Santavicca2010nl} Santavicca, D.F. , Chudow, J.D.,  Prober, D.E. , Purewal, M.S.,
   Kim,  P., \emph{Nano Lett.} \textbf{10}, 4538 (2010).
   \textit{Energy loss of the electron system in individual single-Wwlled carbon nanotubes}

\bibitem{Voisin2015jpcm} Voisin, C., Pla\c{c}ais, B., \emph{J. Phys.: Condens. Matter}
    \textbf{27}, 60301 (2015). \textit{Hot carriers in graphene}

\bibitem{Brunel2015jpcm}
 Brunel, D., et al., \emph{J. Phys.: Condens. Matter} \textbf{27}, 164208 (2015).
\textit{Onset of optical-phonon cooling in multilayer graphene revealed by RF noise and black-body radiation thermometries}

\bibitem{Betz2012prl} Betz, A.C., et al.,  \emph{Phys. Rev. Lett.}
    \textbf{109}, 056805 (2012). \textit{Hot electron cooling by acoustic phonons in graphene}.

\bibitem{McKitterick2016prb} McKitterick, C.B., Prober, D.E., Rooks, M. J.,  \emph{Phys. Rev. B}
    \textbf{93}, 075410 (2016).
    \textit{Electron-phonon cooling in large monolayer graphene devices}

\bibitem{Fong2013prx} Fong, K.C. , et al., \emph{Phys. Rev. X} \textbf{3}, 041008 (2013).
    \textit{Measurement of the electronic thermal conductance channels and heat capacity of graphene at low temperature}

\bibitem{Crossno2016science} Crossno, J., et al., \emph{Science} \textbf{351},
    6277 (2016). \textit{Observation of the Dirac fluid and the breakdown of the Wiedemann-Franz law in graphene}

\bibitem{Bistritzer2009} Bistritzer,  R., MacDonald, A.H., \emph{Phys. Rev. Lett.} \textbf{102},
    206410 (2009). \textit{Electronic cooling in graphene}.

\bibitem{Yao2000prl} Yao, Z.,  Kane, C.L.,  Dekker, C.,  \emph{Phys. Rev. Lett.} \textbf{84}, 2941
    (2000).
 \textit{High-Field electrical transport in single-wall carbon nanotubes}

 \bibitem{Bourlon2004prl}  Bourlon, B., et al., \emph{Phys. Rev. Lett.} \textbf{92}, 026804 (2004).
 \textit{Geometrical dependence of high-bias current in multiwalled carbon nanotubes}

\bibitem{Biehs2010} Biehs, S-A., Rousseau, E., Greffet, J-J., \emph{Phys. Rev. Lett.} \textbf{105},
    234301 (2010). \textit{Electronic cooling in graphene}.

\bibitem{Brida2013ncomm} Brida, D., et al., \emph{Nature Commun.} \textbf{4},
    1987, (2013). \textit{Ultrafast collinear scattering and carrier multiplication in graphene}

\bibitem{Kadi2015scirep} Kadi, F., Winzer, T. Knorr, A., Malic, E. \emph{Scientific Reports}
    \textbf{5}, 16841 (2015).
    \textit{Impact of doping on the carrier dynamics in graphene}

\bibitem{Tielrooij2017arXiv} Tielrooij, K.J. , et al., \emph{arXiv:}1702.03766v1 (2017).
 \textit{ Out-of-plane heat transfer in van der Waals stacks: electron-hyperbolic phonon coupling}

 \bibitem{Dean2010nnano} Dean, C.R., et al., \emph{Nat. Nanotech.} \textbf{5}, 722 (2010).
 \textit{Boron nitride substrates for high-quality graphene electronics}


\end{thebibliography}
\end{document}